\documentstyle[11pt,newpasp,twoside,epsf]{article}

\begin{document}

\title{Evolution of the correlation function 
as traced by the HDF galaxies}

\author{Boudewijn F. Roukema$^{1,2,3}$}
\affil{$^1$Nicolaus Copernicus
Astronomical Center, ul. Bartycka 18, 00-716 Warsaw, Poland\\
$^2$Institut d'Astrophysique de Paris, 98bis Bd Arago,
F-75.014 Paris, France\\
$^3$Inter-University Centre for Astronomy and Astrophysics, 
Post Bag 4, Ganeshkhind, Pune, 411 007, India 
{\em (boud@iucaa.ernet.in)} }

\author{David Valls-Gabaud$^{4,5}$}
\affil{$^4$UMR CNRS 7550, Observatoire de Strasbourg, 
 11 rue de l'Universit\'e, 67000 Strasbourg, France 
{\em (dvg@ast.obs-mip.fr)}\\
$^5$Royal Greenwich Observatory, Madingley Road,
                 Cambridge CB3 0EZ, UK}


%
%



\begin{abstract}
An initially highly biased value of the spatial correlation function,
which decreases with time up to a transition redshift $z_t \sim 1-2$, was
predicted theoretically at least as early as 1993, and shown to be
consistent with the HDF-N angular correlation function estimates in
1997.
 Observational analyses are presented here which show
(i) an HDF-N estimate of the correlation function amplitude of galaxies
selected at $z \sim 2$ (by the UV drop-{\em in} technique), which supports the
estimate $z_t \sim 2$ and 
(ii) an HDF-N estimate of the correlation function
amplitude of galaxies selected at $z \sim 3.7$ by using photometric
redshifts, which suggests that the correlation function amplitude
evolves as $(1+z)^{2\pm3.5}$ during epochs earlier than $z_t$.
\end{abstract}


\keywords{cosmology: theory---galaxies: formation---galaxies: 
clusters: general---galaxies: distances and redshifts---cosmology: 
observations}


\section{Introduction}

The theoretical possibility of an initially highly biased 
spatial two-point auto-correlation function  
of dark matter haloes, $r_{\mbox{\rm \small halo}}$,  which
decreases in amplitude as fluctuations in low-density regions successively 
become non-linear and collapse 
(decreasing correlation period, hereafter DCP),
was noticed at least as early as 1993 (Roukema 1993; 
Brainerd \& Villumsen 1994).
Yamashita (unpublished) also noticed the same effect in 
hydrodynamical N-body simulations, so the effect could follow through
to the galaxy correlation function $\xi$
either for the simplest possible star formation hypotheses 
(e.g. every halo becomes luminous immediately following collapse),
or for less simplistic models.

Because implicit estimates of $\xi$ at high redshift via 
angular correlation function measurements from photometric surveys
were {\em lower} than expected (e.g. Roukema \& Peterson 1994 and references 
therein) in the early 1990's, it seemed that the DCP would have
contradicted the observations. However, 
Ogawa, Roukema \& Yamashita (1997) showed that the HDF-N estimates of the angular 
correlation function by Villumsen (1997) were {\em not} in contradiction
with the DCP, i.e. that HDF observations were consistent with the DCP.

Since then, high redshift ($z\,\lower.6ex\hbox{$\buildrel >\over \sim$} \, 1$)
 galaxy spectroscopy via 
Lyman-break selection and photometric redshift techniques applied
(in particular) to the HDF-N (and -S) have created
a new era in measurement of galaxy statistics. It quickly became clear
that the Lyman-break galaxies, at $z\sim3$, are highly clustered
(Giavalisco et al{.} 1998) and that the DCP was no longer a mere theoretical
prediction. 

Several papers further developing the theory of the DCP 
(Mo \& White 1996; Bagla 1998; Moscardini et al{.} 1998
and references therein) 
have since appeared,
and several observational estimates at 
$z \,\lower.6ex\hbox{$\buildrel >\over \sim$} \,  2$ have been 
made and compared to various theoretical predictions 
(Miralles, Pell\'o \& Roukema 1999; Arnouts et al{.} 1999; 
Magliocchetti \& Maddox 1999).

In parallel with the various model dependent methods of 
analysing the observational results, it is suggested that 
Ogawa et al{.}'s (1997) 
extension of Groth \& Peebles' (1977) power law model of
correlation function evolution via a {\em transition redshift 
$z_t$} and a power law of $(1+z)$ for $z\ge z_t$ should provide
a simple way to characterise and compare different observational
and theoretical analyses. This is presented in \S\ref{s-orylaw}

A complementary observational analysis to the above is that
of estimating $\xi(z \sim 2),$ which is done by a method which is
itself also complementary to the above: the Lyman-break technique
is used to select UV drop-{\em in} galaxies as opposed to 
UV drop-{\em out} galaxies, i.e. those for which $z \,\lower.6ex\hbox{$\buildrel <\over \sim$} \, 
 2.5$.
This analysis is summarised in \S\ref{s-uvdropin}, 
and was carried out using integral constraint corrections
{\em without} reintroducing linear uncertainty 
terms which formulae like that of Landy \& Szalay (1993) are designed to
avoid or minimise (\S\ref{s-nolinear}).

A high bias was {\em not} detected at $z\sim2.$ 
Combination of the Giavalisco et al{.} (1998) 
($z_{\mbox{\rm \small med}}$ $\approx 3$) 
and Miralles et al{.} (1999)
($z_{\mbox{\rm \small med}}$ $ \approx 3.7$) estimates for $\xi$
can then be used to estimate $z_t$ and $\nu$ in eq.~(\ref{e-eps_nu}).
This is presented in \S\ref{s-dcpest}
 
\section{Characterising the DCP} \label{s-orylaw}
After Ogawa et al{.} (1997), an extension of Groth \& Peebles' (1977) 
classical 
formula [Eq.~(\ref{e-eps_nu}) with $z_t \gg 1$] 
provides a way to represent the observational results without
depending on particular galaxy formation models:
\begin{equation}
\xi(r,z)=\left\{ 
        \begin{array}{ll}
        \left[(1+z)\over(1+z_t)\right]^\nu \xi(r,z_t),
                & z > z_t \\
        (r_0/r)^\gamma (1+z)^{-(3+\epsilon-\gamma)}, & z_t \ge z >0
        \end{array}
        \right.
\label{e-eps_nu}
\end{equation}
where $r$ is the galaxy pair separation; $r$ and $r_0$ are expressed
in comoving units; $\gamma\sim 1.8 $ is determined from
observations; $\epsilon$ represents low redshift correlation growth
and has the value $\epsilon=0$ for clustering which is stable (constant) in
proper units on small scales; $z_t$ is a transition redshift 
from the DCP at high $z$ to the low $z$ period of 
correlation growth; and $\nu$ represents 
the rate of correlation decrease at high $z$ [no relation to $\nu$
of Bagla (1998)].

\section{The UV drop-in technique: selecting a $z \sim 2$ sample} 
\label{s-uvdropin}

The UV drop-out technique selects galaxies above $z\approx 2.5$.
In contrast, by selecting only those HDF galaxies for which a
source is detected in the $U$ (F300W) band, Mobasher \& Mazzei (1999)
defined a UV drop-{\em in} sample for which $z\approx 2.5$ was
a strong upper limit in redshift. 
Roukema et al{.} (1999) used Mobasher \& Mazzei's 
photometric redshift estimations to create a subsample in 
the range $1.5 \,\lower.6ex\hbox{$\buildrel <\over \sim$} \, 
 z \,\lower.6ex\hbox{$\buildrel <\over \sim$} \, 
 2.5$.

Roukema et al{.} (1999) estimated $\xi(z\sim2)$ from this sample, finding
that for stable clustering in proper coordinates, 
$r_0\sim2.6^{+1.1}_{-1.7}${\mbox{h$^{-1}$ Mpc}} 
for curvature parameters $\Omega_0=1,\lambda_0=0$,  
or $r_0\sim5.8^{+2.4}_{-3.9}${\mbox{h$^{-1}$ Mpc}} for
$\Omega_0=0.1,\lambda_0=0.9,$ if one does not apply any correction
for effects of the non-zero size of galaxy haloes on $\xi$.
The correction for possible effects of non-zero halo size is discussed in 
Roukema (1999).

{ 
\begin{figure}
\centering 
\centerline{\epsfxsize=10cm
{\epsfbox[26 5 600 470]{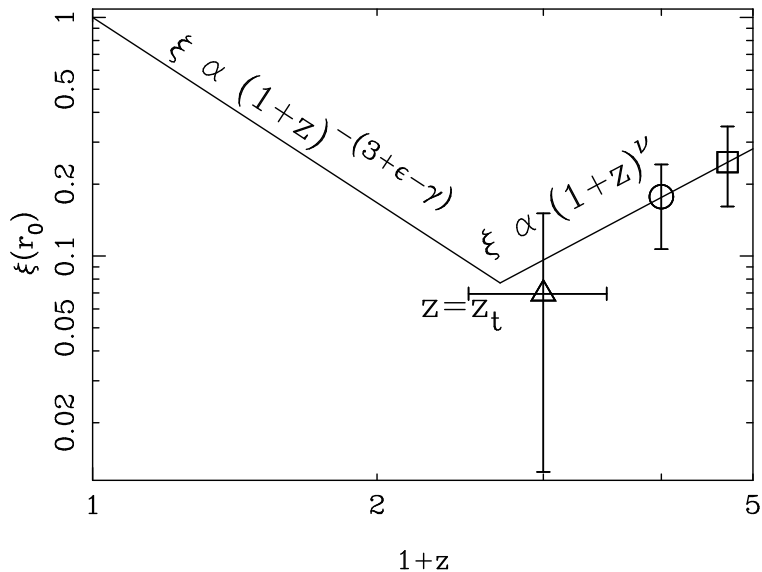} }
}
\caption[]{Spatial correlation function evolution as represented
in Eq.~(\protect\ref{e-eps_nu}), 
with $r_0=5.5$\mbox{h$^{-1}$ Mpc}, and showing observational estimates
of Roukema et al{.} (1999) (triangle), 
Giavalisco et al{.} (1998) (circle) and Miralles et al{.} (1999) (square),
and with $\gamma=1.8,$ $\epsilon=1.4,$ $z_t=1.7$ and $\nu=2.1$,
for $\Omega_0=1, \lambda_0=0$.
\label{f-dcpfig}
}
\end{figure} 
} 

\section{How not to reintroduce linear terms when correcting for 
the integral constraint} \label{s-nolinear}

Correlation function estimates in small fields require integral
constraint corrections. See \S3.1.3 of Roukema et al{.} (1999) 
for a discussion
and references, in particular Hamilton (1993)
for an in-depth 
analysis.

The following formula from Landy \& Szalay (1993):
\begin{equation}
\label{e-ls93ic}
w(\theta) = { N_{gg} - 2 N_{gr} + N_{rr} 
	\over N_{rr} } + C
\end{equation}
where $N_{gg},$ $N_{gr}$ and $N_{rr}$ are numbers of galaxy-galaxy, 
galaxy-random and random-random pairs and $C=0$, avoids linear
terms in the uncertainty of the estimate of $w$ (angular correlation 
function). 

However, it follows from Hamilton (1993) that if $C$ is allowed to 
be a free parameter which is varied in order to match $w(\theta)$
values to a prior hypothesis, e.g. that $w$ is a power law of a given
slope, then linear terms are reintroduced.

Without changing the prior hypothesis,
the way to avoid reintroducing these terms
is to use eq.~(24) of Hamilton (1993):
\begin{equation}
\label{e-wham24}
w(\theta) = { N_{gg} - 2 \overline{n_{\mbox{\small est}}} N_{gr} + 
\overline{n_{\mbox{\small est}}}^2 N_{rr} 
	\over \overline{n_{\mbox{\small est}}}^2 N_{rr} }
\end{equation}
where $\overline{n_{\mbox{\small est}}}$ is 
the mean number density, in principle 
estimated by some means external to the sample, divided by the
number density of the sample itself. 
By treating $\overline{n_{\mbox{\small est}}}$ as
a free parameter instead of $C$, the correction is applied optimally.

\section{DCP parameter estimates} \label{s-dcpest}

From Giavalisco et al{.}'s (1998) estimate 
$r_0=5.3^{+1.0}_{-1.3}${\mbox{h$^{-1}$ Mpc}} ($\Omega_0=1, \lambda_0=0$) 
at $z_{\mbox{\rm \small med}}$ $\approx 3$  
and Miralles et al{.}'s (1999) estimate 
$r_0=7.1\pm1.5${\mbox{h$^{-1}$ Mpc}}
at $z_{\mbox{\rm \small med}}$ $ \approx 3.7$ 
($\Omega_0=1, \lambda_0=0$), the parameters $z_t$ and $\nu$ can be 
estimated from Eq.~(\ref{e-eps_nu}). These are $z_t = 1.7\pm0.9$
and $\nu=2.1\pm3.6$, and are illustrated in
Fig.~1. These values are similar to those expected from
simulations [\S3, \S6 of Ogawa et al{.} (1997); 
also Fig.~3 of Bagla (1998)],
and consistent with the $\xi(z\sim2)$ estimate which indicates that the
DCP has ended by about this epoch.

\section{Conclusion}

The difficulty in estimating photometric redshifts 
at $z\sim2$ can be at least partly overcome by 
applying the UV drop-{\em in} technique. This enables 
studies of galaxy properties 
at an epoch which appears to be an effective transition epoch between 
two periods or regimes of galaxy formation, characterised by a 
minimum in the amplitude of the spatial correlation function.

Further work at this epoch such as that of Martin\'ez et al{.} (1999)
may therefore provide important clues in understanding 
galaxy formation.

\acknowledgments

This research has been supported by the 
Polish Council for Scientific Research Grant
KBN 2 P03D 008 13 and has benefited from 
the Programme jumelage 16 astronomie 
France/Pologne (CNRS/PAN) of the Minist\`ere de la recherche et
de la technologie (France).

\clearpage
\begin{question}{Judy Cohen}
The HDF field subtends a very small angle, so that there are only a
small number of galaxies in the field. Doesn't this {\em scare} you?
\end{question}
\begin{answer}{B.~F.~R.}
The serious violations of international humanitarian law 
allegedly carried out in Yugoslavia by the most 
powerful military coalition on the planet scare me 
(Rangwala et al{.} 1999).
In contrast, for galaxy two-point auto-correlation function estimates, 
the conservative use of error bars 
(e.g. Roukema \& Peterson 1994; Roukema et al{.} 1999)
should help avoid undue emotion. The error
bars on $r_0$, $z_t$ and $\nu$ 
are there for a reason, not just for amusement.
\end{answer}

\end{document}